\documentclass{sigchi-ext}
\usepackage[T1]{fontenc}
\usepackage{textcomp}
\usepackage[scaled=.92]{helvet} %
\usepackage{graphicx} %
\usepackage{balance}  %
\usepackage{booktabs} %
\usepackage{ccicons}  %
\usepackage{ragged2e} %
\usepackage{subcaption}
\usepackage{calc}
\usepackage{marvosym}
\usepackage{adjustbox}
\usepackage{enumitem,kantlipsum}

\definecolor{hotpink}{RGB}{255, 105, 180}

\newcommand{\tthe}{the }
\newcommand{\TThe}{The }

\def\plaintitle{SIGCHI Extended Abstracts Sample File: Note Initial
  Caps} 
\def\emptyauthor{}
\def\plainkeywords{Interactive visualization; deep learning education; machine learning education.}

\title{CNN 101: Interactive Visual Learning for Convolutional Neural Networks}

\numberofauthors{6}
\author{%
  \alignauthor{%
    \textbf{Zijie J. Wang}\\
    \affaddr{Georgia Institute of Technology} \\
    \email{jayw@gatech.edu} }
  \alignauthor{%
    \textbf{Nilaksh Das}\\
    \affaddr{Georgia Institute of Technology}\\
    \email{nilakshdas@gatech.edu} } \vfil 
  \alignauthor{%
    \textbf{Robert Turko}\\
    \affaddr{Georgia Institute of Technology}\\
    \email{rturko3@gatech.edu} }
  \alignauthor{%
    \textbf{Fred Hohman}\\
    \affaddr{Georgia Institute of Technology}\\
    \email{fredhohman@gatech.edu} } \vfil
  \alignauthor{%
    \textbf{Omar Shaikh}\\    
    \affaddr{Georgia Institute of Technology}\\
    \email{oshaikh@gatech.edu} }
  \alignauthor{%
    \textbf{Minsuk Kahng}\\
    \affaddr{Oregon State University}\\
    \email{minsuk.kahng@oregonstate.edu} } \vfil
  \alignauthor{%
    \textbf{Haekyu Park}\\
    \affaddr{Georgia Institute of Technology}\\
    \email{haekyu@gatech.edu} } 
  \alignauthor{%
    \textbf{Duen Horng (Polo) Chau}\\
    \affaddr{Georgia Institute of Technology}\\
    \email{polo@gatech.edu} } 
}

\definecolor{linkColor}{RGB}{6,125,233}
\hypersetup{%
  pdftitle={\plaintitle},
  pdfauthor={\emptyauthor},
  pdfkeywords={\plainkeywords},
  bookmarksnumbered,
  pdfstartview={FitH},
  colorlinks,
  citecolor=black,
  filecolor=black,
  linkcolor=black,
  urlcolor=linkColor,
  breaklinks=true,
}

\copyrightinfo{\scriptsize Permission to make digital or hard copies of part or all of this work for personal or classroom use is granted without fee provided that copies are not made or distributed for profit or commercial advantage and that copies bear this notice and the full citation on the first page. Copyrights for third-party components of this work must be honored. For all other uses, contact the owner/author(s). \\
{\emph{CHI '20 Extended Abstracts, April 25--30, 2020, Honolulu, HI, USA.}} \\
\copyright~2020 Copyright is held by the author/owner(s). \\ 
ACM ISBN 978-1-4503-6819-3/20/04. \\
http://dx.doi.org/10.1145/3334480.3382899}
\begin{document}

\newlength{\skiplength}
\setlength{\skiplength}{\marginparwidth+\marginparsep}

\maketitle

\RaggedRight{} 

\begin{abstract}
The success of deep learning solving previously-thought hard problems has inspired many non-experts to learn and understand this exciting technology.
However, it is often challenging for learners to take the first steps due to the complexity of deep learning models.
We present our ongoing work, CNN 101, an interactive visualization system for explaining and teaching convolutional neural networks.
Through tightly integrated interactive views,
CNN 101 offers both overview and detailed descriptions of how a model works.
Built using modern web technologies, CNN 101 runs locally in users' web browsers without requiring specialized hardware, broadening the public's education access to modern deep learning techniques.
\end{abstract}



\section{Introduction}

Deep learning has become a driving-force in our daily technologies.
Its continued success and potential in various hard problems have attracted immense interest from non-experts to learn this technology.
However, it has a steep learning curve for many beginners.

Since deep learning models are complex, it can be challenging for non-experts to learn the fundamentals.
Inspired by human's brain structure, deep neural network models typically leverage many layers of operations to reach a final computed decision \cite{lecun_deep_2015}.
There are many types of network layers, each having a different structure and underlying mathematical operations.
Therefore, understanding deep learning models requires users to keep track of both \textit{low-level mathematical operations} and \textit{high-level integration} of such operations within the network.

To address this challenge, we are developing \textbf{CNN 101} (\autoref{fig:overview}): an interactive visualization system that helps students learn \textit{convolutional neural networks} (CNN), a foundational deep learning model architecture \cite{lecun_deep_2015}, more easily.
CNN 101 joins the growing body of research that aims to explain the complex mechanisms of modern machine learning algorithms with interactive visualization, such as TensorFlow Playground \cite{smilkov_direct-manipulation_2017} and GAN Lab \cite{kahng_gan_2019}.
For a demo video of CNN 101, visit \url{https://youtu.be/g082-zitM7s}.
In this ongoing work, our primary contributions are:

\begin{enumerate}[topsep=0mm, itemsep=0mm, parsep=1mm, leftmargin=5mm]

\item \textbf{CNN 101}, 
a novel web-based interactive visualization tool that helps users better understand both CNNs high-level model structure and low-level mathematical operations.
Advancing on few existing and prior interactive visualization tools that aim to explain CNN to beginners \cite{harley_interactive_2015, karpathy_convnetjs_2016}, CNN 101 integrates a more practical model and dataset for learners to explore.
Conventionally, deploying deep learning models requires significant computing resources, e.g., servers with powerful hardware.
However, even with a dedicated backend server, it is challenging to support a large number of concurrent users.
Instead, CNN 101 is developed using modern web technologies, where all results are directly computed in users' web browsers.
CNN 101 helps broaden public's education access to modern deep learning technologies.

\item \textbf{Novel interactive visualization design of CNN 101}, which
uses \textit{overview + detail}, \textit{interaction}, and \textit{animation} that simultaneously summarizes complex model structure, and provides context for users to interpret detailed mathematical operations.
CNN 101 presents significant advancement over existing work by explaining how CNNs work at different abstraction levels 
while helping users fluidly transition between such levels to gain a more comprehensive understanding.
Existing work focused on fewer aspects.
For example, Harley et al. \cite{harley_interactive_2015} used a 3D interactive node-link diagram to illustrate CNN structure and neuron activations of a pretrained model, 
but the interface did not visually dissect different neuron's computation processes.
Conversely, expert-facing deep learning visualization tools focus on interpreting what CNN models have learned rather than explaining the underlying operations \cite{hohman_visual_2019}.

\end{enumerate}

We hope our design will inspire research and development of interactive education tools that help democratize more artificial intelligent technologies. %
\newcommand{\vimage}[1]{$\vcenter{\hbox{\includegraphics[height=9pt]{#1}}}$}

\begin{figure*}
\begin{minipage}[t]{0.925\marginparwidth}
  \hspace*{-\skiplength}
  \parbox{0.925\marginparwidth}{
    
    \TThe \textit{Overview} \vimage{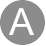} visualizes activation maps of all neurons as heatmaps connected with edges.

    \vspace{1pc}
    When user clicks a convolutional neuron in \vimage{figures/symbol-a.pdf}, the view transitions to \tthe \textit{Convolutional Intermediate View}\\ (\vimage{figures/symbol-a.pdf} \MVRightarrow \hspace{0pt} \vimage{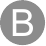}).
    
    \vspace{1pc}
    \TThe \textit{Flatten Intermediate View} appears when an output neuron is selected instead (\vimage{figures/symbol-a.pdf} \MVRightarrow \hspace{0pt} \vimage{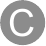}).
    
    \vspace{3pc}
    \vimage{figures/symbol-b.pdf} demonstrates the relationship between selected convolutional neuron and its previous layer.

    \vspace{1pc}
    \vimage{figures/symbol-b.pdf} transitions to \tthe \textit{Detail View} which illustrates the convolution operation on selected input neuron\\ (\vimage{figures/symbol-b.pdf} \MVRightarrow \hspace{0pt} \vimage{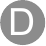}).

    \vspace{1pc}
    \vimage{figures/symbol-c.pdf} explains the flatten layer between the second last layer and output layer.
  }
\end{minipage}%
\begin{minipage}[t]{\textwidth}
  \centering
  \hspace*{-1.85\marginparwidth}
  \parbox{\textwidth}{
    \includegraphics[width=\textwidth]{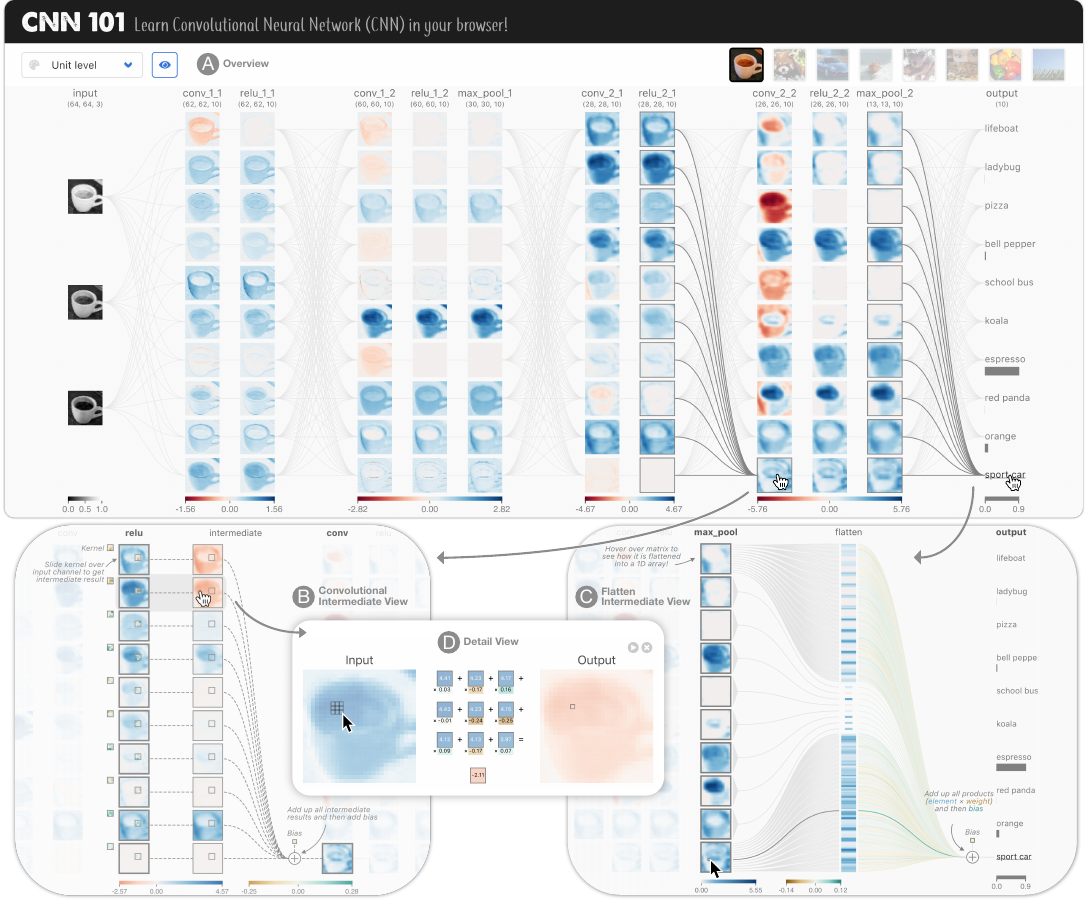}
    \caption{The CNN 101 user interface and its tightly coupled, multiple views.}~\label{fig:overview}
  }
\end{minipage}
\end{figure*}

\section{System Design and Implementation}
CNN 101 is an interactive system for illustrating how a trained CNN model classifies an image (\autoref{fig:overview}).
It enables users to explore the CNN structure and underlying operations in a browser.
To elucidate the CNN's complex process of classifying images, CNN 101 consists of three views: 
(1) \textbf{Overview} (\autoref{fig:overview}A) shows the big picture of the CNN, describing how the input image is connected to the classification likelihood through different layers;\\
(2) \textbf{Intermediate View} (\autoref{fig:overview}B-C) dissects the relationship between one neuron and its previous layer;
(3) \textbf{Detail View} (\autoref{fig:overview}D) interactively visualizes the inner workings of different CNN operations.
Transitions of these views follow an \textit{overview-to-detail} order and are animated to help users assimilate the relationship between different states. 

\begin{marginfigure}[0pc]
  \begin{minipage}{\marginparwidth}
    \centering
    \includegraphics[width=0.9\marginparwidth]{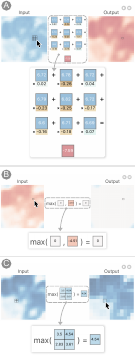}
    \caption{\TThe Detail View explains the underlying mathematical operations of a CNN.
    \protect\vimage{figures/symbol-a.pdf} shows the element-wise dot-product occurring in a convolutional neuron.
    \protect\vimage{figures/symbol-b.pdf} visualizes the Rectified Linear Unit (ReLU) operation, whereas \protect\vimage{figures/symbol-c.pdf} illustrates how max pooling works.
    Users can hover over heatmaps to display an operation's input to output mappings. }~\label{fig:detail}
  \end{minipage}
\end{marginfigure}

\textbf{Overview. }This view is the starting view of CNN 101 (\autoref{fig:overview}A).
It shows activation heatmaps of neurons in all layers.
Neurons in consecutive layers are connected with edges, and hovering over one neuron highlights its incoming edges.
Convolutional and output neurons connect to all neurons in the previous layer, whereas other neurons connect to only one neuron from the earlier layer.

\begin{marginfigure}[-9pc]
  \begin{minipage}{\marginparwidth}
    \centering
    \includegraphics[width=0.91\marginparwidth]{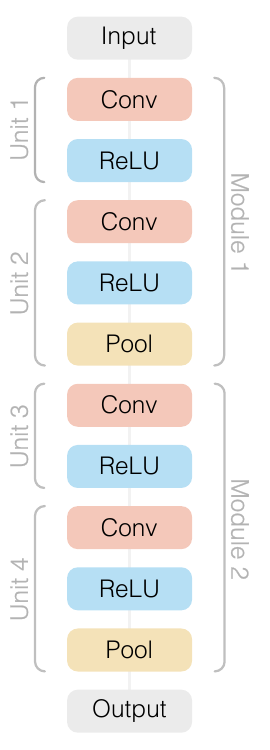}
    \caption{Tiny VGG has the same 3$\times$3 convolution and 2$\times$2 pooling layers as in VGGNet, but with a smaller network depth. Since Tiny VGG layers are grouped into units and modules, CNN 101 allows users to control how each unit and module is colored in the Overview heatmap.
    }~\label{fig:tiny-vgg}
  \end{minipage}
\end{marginfigure}

We show heatmaps with a symmetric diverging red-to-blue colormap where zero is encoded as white.
For example, darker red pixels indicate smaller negative values while darker blue pixels indicate larger positive values.
We group our CNN layers into four units and two modules (\autoref{fig:tiny-vgg}).
Each unit has at most one convolutional layer.
The last two units (Module 1) are duplicate of the first two units (Module 2).
Users can change the heatmap colormap scope based on defined layer groups.
This option enables users to compare neuron activations in different levels and contexts.

\textbf{Intermediate View. }
CNN 101 has two types of Intermediate Views: \tthe Convolutional Intermediate View and \tthe Flatten Intermediate View.
When users click a convolutional neuron in \tthe Overview, \tthe Convolutional Intermediate View (\autoref{fig:overview}B) applies a convolution on each input node of the selected neuron.
Then, it displays these intermediate results as heatmaps.
This view also visualizes associated convolution kernel weights as small heatmaps, which slide over input and intermediate result heatmaps.
This animation mimics the CNN's internal sliding window.
In addition, Edges in \tthe Convolutional Intermediate View are animated as flowing dash-lines, which help signify the order and direction of this intermediate operation.

\TThe Flatten Intermediate View (\autoref{fig:overview}C) explains a flatten layer, which is often used in a CNN to reshape the second last layer into a dense layer, so the fully connected output layer can make classification decisions.
This view encodes each flatten layer neuron as a short line, with the same color as its source element (pixel) in previous layer.
Also, each short line is connected to its source and intermediate result with edges, whose color further encodes model weight value.
When users hovers over an element in the source heatmap, its associated short line  and edges are highlighted.

\textbf{Detail View. }
This view has three variants designed for convolutional (\autoref{fig:detail}A), activation (\autoref{fig:detail}B), and pooling layers (\autoref{fig:detail}C), respectively.
\TThe Detail View provides the user with a low-level, interactive analysis of the mathematical operations occurring at each layer.
Users not only can observe each operation run on an interval displayed by a sliding input region, but also directly interact with Detail View by hovering over pixels to visualize the operation on the selected input region to yield the resulting output values.
By providing a straightforward and interactive visualization of the input and output of multiple fundamental CNN operations, \tthe Detail View allows users who are unfamiliar with CNN mechanisms to understand its mathematical intricacies.

Moreover, with CNN 101's \textit{overview-to-detail} transition hierarchy and \textit{focus + context} layout, users can learn how each single low-level operation contributes to the high-level CNN flow.
For example, a particular convolution output, explained in \tthe Detail View, is only an intermediate result.
To compute the output of a convolutional neuron, one needs to add up all these intermediate results with bias, as described in \tthe Overview and \tthe Intermediate View.
Advancing over existing tools, CNN 101's hierarchical design builds up user's mental model to understand this connection. 

\textbf{CNN Model Training. }
Inspired by the popular deep learning architecture VGGNet \cite{simonyan_very_2015} and Stanford's lecture notes on CNN \cite{karpathy_cs231n_2016}, we train a Tiny VGG on Tiny ImageNet dataset \cite{noauthor_tiny_2015} for demonstration purpose.
Tiny ImageNet has 200 image classes, a training dataset of 100,000 64$\times$64 color images, and a validation/test dataset of 10,000 images each.
Our model is trained using \textit{TensorFlow} \cite{abadi_tensorflow:_2016} on images from 10 selected everyday classes (lifeboat, ladybug, pizza, bell pepper, school bus, koala, espresso, red panda, orange, and sport car) with batch size and learning rate fine-tuned using a 5-fold cross-validation scheme; it achieves a 70.8\% top-1 accuracy on the test dataset.

\textbf{Front-end Visualization. }
We use \textit{TensorFlow.js} \cite{smilkov_tensorflow.js:_2019} to load our trained Tiny VGG and compute forward propagation results directly in user's browser.
We use \textit{D3.js} \cite{bostock_d_2011} to visualize the network structure and implement interactions and animations.
Our implementation is robust, so the this visualization prototpye can be quickly applied to other dataset and linear CNN models.

\section{Preliminary Results: Usage Scenarios}

We now present two usage scenarios where CNN 101 assists users to learn CNN process and gain develop learning intuitions.

\textbf{Understanding layer relationship through visualizing intermediate operations. }
An undergraduate student Sally is learning about various types of CNN layers in her introductory machine learning course.
She does not fully understand how the final output layer maps previous 2D matrices into a class probability number.
Sally starts investigating Tiny VGG with CNN 101 by inspecting layer dimensions.
She quickly noticed that the output layer has dimension 10, while its previous layer has dimension 13$\times$13$\times$10.
Sally hovers over the output class \texttt{sport car} and sees its incoming edges from the previous layer.
CNN 101 helps her quickly recognize essential basic information that there are 10 image classes, 10 neurons in \tthe \texttt{max\_pool\_2} layer, and that each output class connects to 10 previous neurons.
Then, Sally clicks on \texttt{sport car}, causing \tthe \textit{Overview} to transition to \tthe \textit{Flatten Intermediate View}, displaying the flatten layer between \tthe \texttt{max\_pool\_2} layer and \tthe \texttt{sport car} class label (\autoref{fig:overview}C).
By hovering over the heatmap in \tthe \texttt{max\_pool\_2} layer, Sally sees the highlighted edges connecting each matrix element first to the flatten layer, and then to \tthe \texttt{sport car} class label.
Through CNN 101's interactive visualization, Sally realizes that the illustrations in most deep learning tutorials have in fact been skipping this important information in CNN, i.e., that: 
(1) there is actually a ``hidden'' layer that unrolls the output of \tthe \texttt{max\_pool\_2} layer into a 1D array, and 
(2) output layer connects to an intermediate flatten layer instead of directly to \tthe \texttt{max\_pool\_2} layer.

\textbf{Learning layer operation in multiple abstraction levels. }
Harry is a biology researcher who has learned about CNNs in an online deep learning course.
Since he plans to train a CNN model for his project, he uses CNN 101 to review the inner workings of different CNN layers.
Harry launches CNN 101 and skims through all layers on \tthe \textit{Overview}.
He has forgotten what exactly \tthe ReLU layer does upon reaching it in the interface.
However, CNN 101 immediately helps him notice that all previous heatmap red pixels disappear in ReLU layers (\autoref{fig:overview}A).
After selecting other input images and having the same observation, Harry guesses that ReLU layers ignore negative values and only propagate positive values.
He clicks on a ReLU neruon, which causes \tthe \textit{Overview} to transition to \tthe \textit{Detail View}.
Seeing ReLU's underlying equation, $\max\left(0, x\right)$, revealed on \tthe \textit{Detail View} (\autoref{fig:detail}B), Harry is very happy that his hypothesis is validated.
By offering both an overview and a detailed explanation of \tthe ReLU activation function, CNN 101 helps Harry understand how ReLU layer works in different abstraction levels. %
\section{Ongoing Work and Conclusion}

\textbf{User customization. }
We are working on extending CN' 101's interactivity to promote user engagement and to explain more CNN concepts.
Besides choosing an input image from Tiny ImageNet, we plan to support users to upload their own images, capture images from webcam, and free form drawing.
These options can enable users to engineer images to test their hypothesis regarding CNN operations during learning \cite{kahng_how_2019}.
For example, if one user is confused about how the convolution operation works on multiple channels, she can create an image that only has non-zero values in the red channel and feed it into CNN 101.
Then she can learn that convolutions are performed independently on input channels, by observing the intermediate results and activation maps on the first convolution layer.

Currently CNN 101 explains convolution, activation, and pooling operations at single-neuron-level as well as layer-level.
However, these operations have fixed hyper-parameters.
For example, the convolution process always uses a 3$\times$3 kernel with padding of 0 and stride of 1.
We are working to support
users to configure these settings and observe the results in real time.
Such interactive ``hypothesis testing'' and experimentation help
users learn other advanced
deep learning architectures more easily.

\textbf{Planned evaluation. }
Despite the increasing popularity of applying interactive visualization to teach deep learning concepts, little work have been done to evaluate how effective these tools are \cite{kahng_how_2019}.
We plan to run a user study to compare the educational effectiveness of CNN 101 and that of conventional educational mediums such as (static) tutorials, textbooks, and YouTube lecture videos.
We plan to recruit undergraduate students who have basic machine learning background and are new to deep learning.
Our study will have two conditions: \textit{CNN 101} v.s. \textit{conventional tools}. We will randomly assign students into these conditions, and they will use respective tools to learn how CNN works.
Each participant will complete a pre-test quiz and a post-test quiz, allowing us to quantify and more deeply understand the education effectiveness of CNN 101.

\textbf{Deployment. }
We are working to deploy and open-source CNN 101, similar to TensorFlow Playground \cite{smilkov_direct-manipulation_2017} and GAN Lab \cite{kahng_gan_2019}, so that it will be easily accessible by learners from all over the world.

\textbf{Conclusion. }
CNN 101 takes steps toward democratizing deep learning that has been closely impacting people's daily lives.
Through applying interactive visualizing techniques, CNN 101 provides users with an easier way to learn deep learning mechanisms and build up neural network intuitions.
We plan to extend CNN 101's capabilities to support further user customization and personalized learning;
we will deploy and open-source CNN 101 and also evaluate it in depth to help build design principles for future deep learning educational tools. 
\section{Acknowledgements}
We thank Anmol Chhabria for helping to collect related interactive visual education tools.
This work was supported in part by NSF grants IIS-1563816, CNS-1704701, NASA NSTRF, DARPA GARD, gifts from Intel (ISTC-ARSA), NVIDIA, Google, Symantec, Yahoo! Labs, eBay, Amazon.

\balance{} 

\bibliographystyle{SIGCHI-Reference-Format}
\bibliography{cnn101}

\end{document}